\documentclass[aps,superscriptaddress,showpacs,nofootinbib,eqsecnum]{revtex4}

\usepackage{amsmath}
\usepackage{mathrsfs}

\usepackage{epsfig}  \input{epsf}
\def\be{\begin{equation}} \def\ee{\end{equation}} \def\bea{\begin{eqnarray}}
\def\eea{\end{eqnarray}}

\def\beq{\begin{equation}}
\def\eeq{\end{equation}}
\def\beqa{\begin{eqnarray}}
\def\eeqa{\end{eqnarray}}

\def\MeV{\rm MeV}
\def\fm{\rm fm}

\pacs{21.65.Qr., 26.60.Kp, 11.10.Wx, 11.30.Rd, 12.39.Ki}
\begin{document}

\title{The Surface Tension of Magnetized Quark Matter}

\author{Andr\'{e} F. Garcia}
\affiliation{Departamento de F\'{\i}sica, Universidade Federal de Santa
  Catarina, 88040-900 Florian\'{o}polis, Santa Catarina, Brazil}

\author{Marcus Benghi Pinto} \email{marcus@fsc.ufsc.br}
\affiliation{Departamento de F\'{\i}sica, Universidade Federal de Santa
  Catarina, 88040-900 Florian\'{o}polis, Santa Catarina, Brazil}

\date{\today}

\begin{abstract}
The surface tension of quark matter plays a crucial role
for the possibility of quark matter nucleation
during the formation of compact stellar objects and also for the existence of a mixed phase within hybrid stars. However, despite its importance, this quantity does not have a well  established numerical value. Some early estimates have predicted that, at zero temperature, the value falls within the wide range $\gamma_0\approx10-300{\rm\ MeV/fm^2}$  but, very recently, different model applications have reduced these numerical values to fall within the  range  $\gamma_0\approx5-30{\rm\ MeV/fm^2}$  which would favor the phase conversion process as well as the appearance of a mixed phase in hybrid stars.
In magnetars one should also account for the presence of very high magnetic fields which may reach up to about $ eB\approx 3-30\, m_\pi^2$ ($B \approx 10^{19}-10^{20} \,G$) at the core of the star so that it may also be important to analyze how the presence of a magnetic field affects the surface tension. With this aim we consider  magnetized two flavor quark matter, described by the Nambu--Jona-Lasinio model. We show  that although the surface tension oscillates around its $B=0$ value,  when $0 < eB \lesssim 10 \, m_\pi^2$, it only reaches values which are  still relatively small.  For $eB \approx 5\, m_\pi^2$ the $B=0$ surface tension value drops by about $30\%$ while for $eB \gtrsim 10\, m_\pi^2$ it quickly raises with the field intensity so that the phase conversion and the presence of a mixed phase should be suppressed if extremely high fields are present. We also investigate how thermal effects influence the surface tension for magnetized quark matter.

\end{abstract}

\maketitle

%========================================================================
\section{Introduction}

The understanding of compact stars requires the study of strongly interacting matter at
low temperatures and high chemical potentials.  However, this portion of the QCD phase diagram cannot be addressed by current lattice-QCD methods so that  studies of this phase region must  rely on less fundamental models.
Most investigations suggest that there is a first-order chiral phase transition
which, for $T\approx0$, sets in at baryon densities several times that of the
nuclear saturation density, $\rho_0\approx0.17 \, {\rm fm}^{-3}$. The expected
phase transition will have significant implications for the possible existence of
quark stars and the possibilities depend on the dynamics of the phase conversion as well as
 on the time scales involved \cite {ed1,bie,sagert,bruno,bombaci}.
%This is because chiral symmetry may be restored already during
%the early post-bounce accretion stage of a core-collapse supernova event
%and the associated neutrino burst might then provide a spectacular signature
%for the presence of quark matter inside compact stars \cite {sagert}.
%However, as pointed out in Refs.\ \cite{bruno,bombaci},
%the possibilities depend on the dynamics of the phase conversion
%and especially on the time scales involved.
When the phase diagram of bulk matter exhibits a first-order phase transition,
the two phases, associated with a high and a low density value ($\rho^H$ and $\rho^L$), may coexist in mutual thermodynamic equilibrium and,
consequently, when brought into physical contact
a mechanically stable interface will develop between them.
The associated surface tension $\gamma_T$
depends on the temperature $T$; it has its largest magnitude at $T=0$
approaching zero as $T$ is increased to the critical end point temperature, $T_c$,
where the first order transition line terminates.
The surface tension plays a key role in the phase conversion process
and it is related to various characteristic quantities such as
the nucleation rate, the critical bubble radius, and the favored scale of the
blobs generated by the spinodal instabilities \cite{PhysRep,RandrupPRC79}.
For our present purposes it is important to remark  that a small surface tension would
facilitate various structures in compact stars,
including the presence of mixed phases in a hybrid star \cite {kurkela}.

Unfortunately, despite its central importance,
the surface tension of quark matter is rather poorly known.
At vanishing temperatures, some early estimates  fall within a wide range,
typically $\gamma_0\approx10-50~\MeV/{\rm fm}^2$
\cite{heiselberg,sato} and values of $\gamma_0\approx30\,\MeV/\fm^2$
have been considered for studying the effect of quark matter
nucleation on the evolution of proto-neutron stars \cite{constanca}.
But the authors in Ref.\ \cite {russo}, taking into
account the effects from charge screening and structured mixed phases,
estimate $\gamma_0\approx50-150~\MeV/{\rm fm}^2$, without excluding smaller values,
and even a higher value, $\gamma_0\approx300~\MeV/{\rm fm}^2$,
was found
on the basis of dimensional analysis of the minimal
interface between a color-flavor locked phase and nuclear matter \cite {alford}.

More recently, the surface tension for two-flavor quark matter was evaluated, in Ref. \cite {leticia}, within
the  the quark meson model (QM), in the framework of the thin-wall approximation for bubble
nucleation . The predicted  values cover the  $5-15\,\MeV/\fm^2$ range,
depending on the inclusion of vacuum and/or thermal corrections.
In principle, this range makes nucleation of quark matter possible
during the early post-bounce stage of core-collapse supernovae
and it is thus a rather important result.

The Nambu--Jona-Lasinio model (NJL) with two and three flavors was subsequently  considered in the evaluation of $\gamma_T$ \cite {mvj} via a geometrical approach  introduced by Randrup  in Ref. \cite {RandrupPRC79}.
This method makes it possible
to express the surface tension for any subcritical temperature in terms of
the free energy density for uniform matter in the unstable density range.
In practice, the  procedure is rather simple to implement
and it provides an estimate for the surface tension
that is consistent with the equation of state (EoS) implied by the adopted model,
with its specific approximations and parametrizations. The results obtained in Ref. \cite {mvj} predict that, at zero temperature, $\gamma_0 \approx 7-30 \, \MeV/\fm^2$ depending on the chosen parameters.

Very recently, the Polyakov quark meson model (PQM) with three flavors has been considered \cite {Mintz} in the context of the thin wall approximation extending the work of Ref. \cite {leticia} with confinement and strangeness. Depending on the adopted parametrization, the numerical results obtained in Ref. \cite {Mintz} are within the  $\gamma_0 \approx 13-28 \, \MeV/\fm^2$ range. The authors have confirmed that the inclusion of the strange sector, which was originally done in Ref. \cite {mvj}, does not change appreciably the dynamics of the transition at low temperatures and high chemical potentials as neither does the inclusion of the Polyakov loop. Regarding the possibility of phase conversion taking place, within compact stellar objects, it is important to remark that all these three recent evaluations \cite {leticia, mvj,Mintz} predict values for the surface tension which are  low enough so that, in principle, the phase conversion phenomenon could take place. At the sam
 e time, these three  estimates, of low values, favor the appearance of a mixed phase within a hybrid star.

One should also recall that  very high magnetic fields can be present in magnetars reaching up to
$ eB\approx 3-30\, m_\pi^2$ ($B \approx10^{19}-10^{20} \,G$), or higher, at the core of the star \cite {magnetars}. In many applications this type of compact stellar objects are modeled as a hybrid star which has a core of quark matter surrounded by hadronic matter \cite {veronica} and if the surface tension between the two phases is small enough, as predicted by Refs. \cite {leticia,mvj,Mintz}, the transition occurs via a  mixed phase (Gibbs construction). On the other hand, if $\gamma_T$ has a high value, as predicted by Refs.  \cite {russo,alford}, it occurs at a sharp interface (Maxwell construction) \cite {mark}. Therefore, the value of the surface tension in the presence of high magnetic fields may be an important ingredient for  investigations related to quark and hybrid stars. Since this type of evaluation does not seem to have been carried out before  we intend to perform such a calculation here by extending the work of Ref. \cite {mvj} so as to account for the presenc
 e of high magnetic fields. The coexistence region associated with the first order transition of strongly interacting magnetized matter has been recently investigated in Ref. \cite {andre} which predicts, as one of its main results, that the value of $\rho^H$  oscillates around the $B=0$ value for  $0< eB \lesssim 6 m_\pi^2$ and then grows for higher values. Taking into account that $\gamma_T$ depends on the difference between $\rho^H$ and $\rho^L$ \cite{PhysRep,RandrupPRC79} one may then expect to find a similar behavior here.
Indeed, as we will demonstrate, when a magnetic field is present the surface tension value  oscillates  very mildly  for $0< eB \lesssim 4 m_\pi^2$ before decreasing in a significant way between
$4 m_\pi^2 \lesssim eB \lesssim 6  m_\pi^2$. Then, after reaching a minimum at $eB \approx 6 m_\pi^2$ it  starts to increase continuously  reaching the $B=0$ value at $eB \approx 9 \; m_\pi^2$ which allows to conclude that the existence of a mixed phase remains possible within this range of magnetic fields. For $eB$ values higher than $\approx 10 m_\pi^2$ this quantity increases rapidly with the magnetic field disfavoring    the presence of a mixed phase within hybrid stars. We also show how the temperature affects $\gamma_T(B)$ by decreasing its value towards zero which is achieved at $T=T_c$, as already emphasized.

The paper is organized as follows.
In the next section we review the method for extracting the surface tension
from the equation of state.
In Sec.\ \ref{models} we  present the EoS for  the magnetized two flavor NJL.
 Then, in Sec.\ \ref{results}, we present our numerical results.
The conclusions and final remarks are presented in Sec.\ \ref{conclude}.

%========================================================================
\section{The Geometric Approach to the Surface Tension Evaluation }
\label{method}

To make  this work self contained let us review, in this section,  the geometric approach to the surface tension evaluation which was originally proposed in Ref. \cite  {RandrupPRC79}.
We first assume that the material at hand, strongly interacting matter,
may appear in two different phases under the same thermodynamic conditions
of temperature $T$, chemical potential $\mu$, and pressure $P$.
These two coexisting phases have different values of other relevant quantities,
such as the energy density $\cal E$, the net quark number density $\rho$,
and the entropy density $s$.
Under such circumstances,
the two phases will develop a mechanically stable interface if placed
in physical contact. An interface tension,  $\gamma_T$, is then associated to this interface.

The two-phase feature appears for all temperatures below the
critical value, $T_c$.  Thus, for any subcritical temperature, $T<T_c$,
hadronic matter at the density $\rho^L(T)$
has the same chemical potential and pressure
as quark matter at the (larger) density $\rho^H(T)$.
As $T$ is increased from zero to $T_c$,
the coexistence phase points $(\rho^L,T)$ and $(\rho^H,T)$ trace out the
lower and higher branches of the phase coexistence boundary, respectively,
gradually approaching each other and finally coinciding for $T=T_c$.
Any $(\rho,T)$ phase point outside of this boundary corresponds to
thermodynamically stable uniform matter,
whereas uniform matter prepared with a density and temperature corresponding
to a phase point inside the phase coexistence boundary
is thermodynamically unstable and prefers to separate into
two coexisting thermodynamically stable phases
separated by a mechanically stable interface.
Because such a two-phase configuration is in global thermodynamic equilibrium,
the local values of $T$, $\mu$, and $P$ remain unchanged
as one moves from the interior of one phase through the interface
region and into the interior of the partner phase,
as the local density $\rho$ increases steadily from
the lower coexistence value $\rho^L$
to the corresponding higher coexistence value $\rho^H$.

It is convenient to work in the canonical framework
where the control parameters are temperature and density.
The basic thermodynamic function is thus $f_T(\rho)$,
the free energy density as a function of the (net) quark number density $\rho$
for the specified temperature $T$.
The chemical potential can then be recovered as
$\mu_T(\rho)=\partial_\rho f_T(\rho)$,
and the entropy density as $s_T(\rho)=-\partial_T f_T(\rho)$,
so the energy density is ${\cal E}_T(\rho)=f_T(\rho)-T\partial_Tf_T(\rho)$,
while the pressure is $P_T(\rho)=\rho\partial_\rho f_T(\rho)-f_T(\rho)$.

For single-phase systems $f_T(\rho)$ is convex,
{\em i.e.}\ its second derivative $\partial_\rho^2f_T(\rho)$ is positive,
 while the appearance of a concavity in $f_T(\rho)$
signals the occurrence of phase coexistence, at that temperature.
This is easily understood because when $f_T(\rho)$ has a local concave
anomaly then there exist a pair of densities, $\rho^L$ and $\rho^H$,
for which the tangents to $f_T(\rho)$ are common.
Therefore $f_T(\rho)$ has the same slope at those two densities,
so the corresponding chemical potentials are equal,
$\mu_T(\rho^L)=\partial_\rho f_T(\rho^L)
	=\partial_\rho f_T(\rho^H)=\mu_T(\rho^H)$.
Furthermore, because a linear extrapolation of $f_T(\rho)$ leads from
one of the touching points to the other, also the two pressures are equal,
$P_T(\rho^L)=\rho^L\partial_\rho f_T(\rho^L)-f_T(\rho^L)
	=\rho^H\partial_\rho f_T(\rho^H)-f_T(\rho^H)=P_T(\rho^H)$.
So uniform matter at the density $\rho^L$ has the same temperature,
chemical potential, and pressure as uniform matter at the density $\rho^H$.
The common tangent between the two coexistence points
corresponds to the familiar Maxwell construction and shall here
be denoted as $f_T^M(\rho)$.
Obviously, $f_T(\rho)$ and $f_T^M(\rho)$ coincide at the two coexistence
densities and, furthermore,
$f_T(\rho)$ exceeds $f_T^M(\rho)$ for intermediate densities.
Therefore we have $\Delta f_T(\rho)\equiv f_T(\rho)-f_T^M(\rho)\geq0$.

For a given (subcritical) temperature $T$,
we now consider a configuration in which the two coexisting bulk phases
are placed in physical contact along a planar interface.
The associated equilibrium profile density is denoted by $\rho_T(z)$
where $z$ denotes the location in the direction normal to the interface.
In the diffuse interface region, the corresponding local free energy density,
$f_T(z)$, differs from what it would be for the corresponding Maxwell system,
{\em i.e.}\ a mathematical mix of the two coexisting bulk phases
with the mixing ratio adjusted to yield an average density
equal to the local value $\rho(z)$.
This local deficit amounts to
\begin{equation}
\delta f_T(z)\ =\  f_T(z)-f_i
	-\frac{f_T(\rho^H)-f_T(\rho^L)}{\rho^H-\rho^L}\,(\rho_T(z)-\rho_i)\ ,
\end{equation}
where $\rho_i$ is either one of the two coexistence densities.
The function $\delta f_T(z)$ is smooth and it tends quickly to zero
away from the interface where $\rho_T(z)$ rapidly approaches $\rho_i$
and $f_T(z)$ rapidly approaches $f_T(\rho_i)$.
The interface tension $\gamma_T$ is the total deficit in free energy
per unit area of planar interface,
\begin{equation}
\gamma_T\ =\ \int_{-\infty}^{+\infty}\delta f_T(z)\,dz\ .
\end{equation}

As discussed in Ref.\ \cite{RandrupPRC79},
when a gradient term used to take account of finite-range effects,
the tension associated with the interface between the two phases
can be expressed without explicit knowledge about the profile functions
but exclusively in terms of the equation of state for uniform
(albeit unstable) matter,
\begin{equation}
 \gamma_T\ =\ a \int_{\rho^L(T)}^{\rho^H(T)}
\left[ 2 {\cal E}^g \Delta f_T(\rho)\right]^{1/2} \frac {d \rho}{\rho^g}\ ,
\label{gamma}
\end{equation}
where $\rho^g$ is a characteristic value of the density
and ${\cal E}^g$ is a characteristic value of the energy density,
while the parameter $a$ is an effective interaction range related to
the strength of the gradient term, $C=a^2 {\cal E}^g/(\rho^g)^2$.
We choose the characteristic phase point to be in the middle of the
coexistence region, $\rho^g=\rho_c$ and
${\cal E}^g=[{\cal E}_0(\rho_c)+{\cal E}_c]/2$,
where ${\cal E}_0(\rho_c)$ is energy density at $(\rho_c,T=0)$,
while ${\cal E}_c$ is energy density at the critical point $(\rho_c,T_c)$.
The length $a$ is a somewhat adjustable parameter
governing the width of the interface region
and the magnitude of the tension \cite{RandrupPRC79}.
In Ref. \cite {mvj} this parameter was set to $a\approx1/m_\sigma\approx0.33\,\fm$
which, also,
is approximately the value found in an application of the Thomas-Fermi
approximation to the NJL model  \cite {russos}.
Therefore, we shall adopt the value $a=0.33\,{\rm fm}$
throughout the present work. With these parameters fixed (see Ref. \cite {mvj}),
the interface tension can be calculated once the free energy density
$f_T(\rho)$ is known for uniform matter in the unstable phase region,
$\rho^L(T)\leq\rho\leq\rho^H(T)$.
%While this is straightforward in a canonical formulation,
%where each $(\rho,T)$ characterizes only one manifestation of the system,
%even inside the unstable phase region,
%the task is more complicated in the commonly used grand canonical formulation
%because a given $(\mu,T)$ phase point characterizes three different
%manifestations of the system,
%one stable, one metastable, and one unstable.
%The metastable solutions are located near the coexistence densities,
%while the unstable solutions are located in the intermediate spinodal region
%where uniform matter is mechanically unstable so that even infinitesimal
%irregularities may be exponentially amplified.
%By contrast, only irregularities of a sufficient amplitude are amplified
%in the metastable regions, leading towards either nucleation
%(near the lower coexistence density $\rho^L$)
%or bubble formation (near the higher coexistence density $\rho^H$).

%========================================================================
\section{The EoS for the Magnetized Two Flavor NJL   Quark Model}
\label{models}

The NJL model is described by a Lagrangian density for fermionic fields given by~\cite{njl}

\begin{equation}
\mathcal{L}_{\rm NJL}={\bar \psi}\left( i{\partial \hbox{$\!\!\!/$}}-m\right) \psi
+G\left[ ({\bar \psi}\psi)^{2}-({\bar{\psi}} \gamma _{5}{\vec{\tau}}\psi
  )^{2}\right] ,
\label{njl2}
\end{equation}
\noindent
where $\psi$ (a sum over flavors and color degrees of freedom is implicit)
represents a flavor iso-doublet ($u,d$ type of quarks) $N_{c}$-plet quark
fields, while $\vec{\tau}$ are isospin Pauli matrices. The Lagrangian density
(\ref{njl2}) is invariant under (global) $U(2)_{\rm f}\times SU(N_{c})$ and,
when $m=0$, the theory is also invariant under chiral $SU(2)_{L}\times
SU(2)_{R}$.
Within the NJL model a sharp cut off ($\Lambda$) is generally used as an ultra violet regulator and since the
model is nonrenormalizable, one has to  fix $\Lambda$ to a value related to the
physical spectrum under investigation. This strategy turns the 3+1 NJL model
into an effective model, where $\Lambda$ is treated as a parameter. Then, the phenomenological values
of quantities such as the pion mass ($m_{\pi}$),  the pion decay constant
$(f_{\pi})$, and the quark condensate ($ \langle {\bar \psi} \psi \rangle$) are used to fix  $G$, $\Lambda$, and
$m$.  Here, we choose the set
 $\Lambda=590\,\rm MeV$ and $G\Lambda^2=2.435$ with $m=6\,\rm MeV$
in order to reproduce $f_\pi=92.6\,\MeV$, $m_\pi= 140.2\,\MeV$, and $ \langle {\bar \psi} \psi \rangle^{1/3}=-241.5 \, \MeV$ \cite {frank}. In the MFA the NJL thermodynamic  potential can  can be written as follows \cite {buballa,prcsu2} (see Ref. \cite {prc1} for results beyond MFA)

\begin{equation}
\Omega^{\rm NJL}=\frac{(M-m)^2}{4G}+\frac{i}{2}{\rm tr} \int \frac{d^4p}{(2\pi)^4} \ln  [-p^2+M^2] \,\,,
\label{free}
\vspace{0.4 cm}
\end{equation}

\noindent where $M$ is the constituent quarks mass. In order to study the effect of a magnetic field in the chiral transition at finite temperature and chemical potential a dimensional reduction is induced  via  the following replacements  in Eq. (\ref{free}) \cite{eduana}

\begin{equation}
p_0\rightarrow i(\omega_{\nu}-i\mu)\,\, , \nonumber
\end{equation}

\begin{equation}
p^2 \rightarrow p_z^2+(2n+1-s)|q_f|B \,\,\,\,\, , \,\,\mbox{with} \,\,\, s=\pm 1 \,\,\, , \,\, n=0,1,2...\,\,, \nonumber
\end{equation}

\begin{equation}
\int_{-\infty}^{+\infty}  \frac{d^4p}{(2\pi)^4}\rightarrow i\frac{T |q_f| B}{2\pi}\sum_{\nu=-\infty}^{\infty}\sum_{n=0}^{\infty}\int_{-\infty}^{+\infty} \frac{dp_z}{2\pi} \,\,,\nonumber
\end{equation}

\noindent where $\omega_\nu=(2\nu+1)\pi T$, with $\nu=0,\pm1,\pm2...$ represents the Matsubara frequencies for fermions, $n$ represents the Landau levels and $|q_f|$ is the absolute value of the quark electric charge ($|q_u|= 2e/3$, $|q_d| = e/3$ with $e = 1/\sqrt{137}$ representing the electron charge\footnote{We use Gaussian natural units where $1\, \MeV^2 = 1.44 \times 10^{13} \,{\it G}$ which sets $m_\pi^2/e \simeq 3\times 10^{18}\, G$.}).  Note also that here we have taken the chemical equilibrium condition by setting $\mu_u=\mu_d=\mu$. Then, following Ref. \cite {prcsu2}, we can write the  free energy as

\begin{equation}
\Omega^{\rm NJL}=\frac{(M-m)^2}{4G} + \Omega_{\rm vac}^{\rm NJL}+\Omega_{\rm mag}^{\rm NJL}+\Omega_{\rm med}^{\rm NJL} \,\,\,\,,
\end{equation}

\noindent where

\begin{equation}
\Omega_{\rm vac}^{\rm NJL}=-2N_cN_f \int \frac{d^3{\bf p}}{(2\pi)^3}({\bf p}^2 +M^2)^{1/2} \,\,\,.
\label{vacnjl}
\end{equation}
This divergent integral is regularized by a sharp cut-off, $\Lambda$, yielding
\begin{equation}
\Omega_{\rm vac}^{\rm NJL}= \frac{N_c N_f}{8\pi^2} \left \{ M^4 \ln \left [
    \frac{(\Lambda+ \epsilon_\Lambda)}{M} \right ]
 - \epsilon_\Lambda \, \Lambda\left[\Lambda^2 +  \epsilon_\Lambda^2 \right ] \right \}\,\,,
\end{equation}
where we have defined $\epsilon_\Lambda=\sqrt{\Lambda^2 + M^2}$.
The magnetic and the in-medium terms are respectively given by
\begin{equation} \label{Vmag}
\Omega_{\rm mag}^{\rm NJL}=-\frac{N_c}{2\pi^2}\sum_{f=u}^d (|q_f|B)^2\left\{\zeta^{(1,0)}(-1,x_f)-\frac{1}{2}[x_f^2-x_f]\ln(x_f)+\frac{x_f^2}{4} \right\}\,\,\,,
\end{equation}
and
\begin{equation} \label{Vmed}
\Omega_{\rm med}^{\rm NJL}=-\frac{N_c}{2\pi}\sum_{f=u}^d\sum_{k=0}^{\infty}\alpha_k |q_f|B \int_{-\infty}^{+\infty} \frac{dp_z}{2\pi}\left\{T \ln [1+{e}^{-[E_{p,\,k}(B)+\mu]/T}]+ T \ln [1+{e}^{-[E_{p,\,k}(B)-\mu]/T}]\right\} \,.
\end{equation}

\noindent In the last equation we have replaced the label $n$ by $k$ in the Landau levels in order to account for the degeneracy factor $\alpha_k=2-\delta_{0k}$. Also, in Eq (\ref{Vmag}) we have used $x_f=M^2/(2|q_f|B)$ and $\zeta^{(1,0)}(-1,x_f)=d\zeta(z,\,x_f)/dz|_{z=-1}$ with $\zeta(z,\,x_f)$ representing  the Riemann-Hurwitz function (the details of  the manipulations leading to the equations above can be found in the appendix of Ref. \cite {prcsu2}). Finally, in Eq. (\ref{Vmed}) we have $E_{p,\,k}(B)=\sqrt{p_z^2+2k|q_f|B +M^2}$ where  $M$  is  the effective
self consistent quark mass

\begin{eqnarray}
M &=& m +\frac{ N_c N_f MG}{\pi^2} \left \{
\Lambda
\sqrt {\Lambda^2 + M^2} -\frac{M^2}{2}
\ln \left [ \frac{(\Lambda+ \sqrt {\Lambda^2 + {M^2}})^2}{{M}^2} \right ] \right \} \nonumber \\
&+&\frac{N_c MG}{\pi^2}\sum_{f=u}^d  |q_f| B\left \{ \ln [\Gamma(x_f)]
-\frac {1}{2} \ln (2\pi) +x_f -\frac{1}{2} \left ( 2 x_f-1 \right )\ln(x_f) \right \} \nonumber \\
&-& \frac{N_cMG}{2\pi^2}\sum_{f=u}^d \sum_{k=0}^{\infty} \alpha_k |q_f|B \int_{-\infty}^{\infty} \frac{dp_z}{E_{p,k}(B)}\left\{ \frac{1}{e^{[E_{p,k}(B)+\mu]/T}+1}+\frac{1}{e^{[E_{p,k}(B)-\mu]/T}+1} \right\} \,\,.
\label{MmuB}
\end{eqnarray}
Note that in principle one should have two coupled gap equations for the two distinct flavors: $M_u = m_{u} - 2G(\langle {\bar u} u \rangle + \langle {\bar d} d \rangle)$ and  $M_d = m_{d} - 2G(\langle {\bar d} d \rangle + \langle {\bar u} u \rangle)$ where $\langle {\bar u} u \rangle$ and  $\langle {\bar d} d \rangle$ represent the quark condensates which differ, due to the different electric charges. However, in the two flavor case, the different condensates contribute to $M_u$ and $M_d$ in a symmetric way and since $m_u=m_d=m$ one has $M_u=M_d=M$.

The minimum value of the grand potential represents minus
the equilibrium pressure, $\Omega_{\rm min}(T,\mu)=-P$,
so the net quark number density is given by $\rho=(\partial P/\partial\mu)_T$.
The entropy density given by $s=(\partial P/\partial T)_{\mu}$,
while the energy density, $\cal E$, can then be obtained by means of
the standard thermodynamic relation $P=Ts-{\cal E}+\mu \rho$. The knowledge of all these quantities allow us to determine
the free energy density, $f\equiv{\cal E}-Ts=\mu\rho-P$, as well as the numerical inputs $\rho^H$, $\rho^L$, $\rho^g$, and $\epsilon^g$ which are needed in the evaluation of the surface tension. As  already emphasized, the numerical value for the length scale $a$ is chosen to be $1/m_\sigma \simeq 0.33\,\fm$
(which is about the value found in a Thomas-Fermi application
to the NJL model \cite{russos}).

%========================================================================
\section{Numerical Results}
\label{results}

Let us start the numerical evaluations by obtaining the phase diagram in the $T-\rho_B$ plane in order  to determine   the values of   essential quantities such as $T_c$, $\mu_c$, $\rho^H$, $\rho^L$ which allow for the evaluation  of the inputs
 $\rho^g$, and ${\cal E}^g$ for each value of $B$.
As it is well known, for given subcritical  temperature in the $T-\rho_B$ plane one observes that the associated density region
is bounded by the two coexistence densities $\rho^L$ and $\rho^H$,
for which the chemical potential $\mu$ has the same value,
as does the pressure $P$.
As the density $\rho$ is increased through the lower mechanically metastable
(nucleation) region, $\mu$ and $P$ rise steadily until the lower spinodal
boundary has been reached.
Then, as $\rho$ moves through the mechanically unstable (spinodal) region,
both $\mu$ and $P$ decrease until the higher spinodal boundary is reached.
They then increase again as $\rho$ moves through the higher mechanically
metastable (bubble-formation) region,
until they finally regain their original values at $\rho=\rho^H$.
Fig. \ref{fig1} displays the coexistence region, in the $T-\rho_B$ plane, for $B=0$, $eB=6 m_\pi^2$, and $eB=15 m_\pi^2$.
Noting that  $\rho^H$ oscillates around the $B=0$ value and recalling that $\gamma_T$ depends on the difference between $\rho^L$ and $\rho^H$, see Eq. (\ref {gamma}), one can then expect that the surface tension value at $eB=6 m_\pi^2$ will be smaller than at $B=0$, at least for small  temperatures. On the contrary, for $eB=15 m_\pi^2$, one may expect $\gamma_T$ to assume values much larger than those obtained in the $B=0$ case. These expectations will be explicitly confirmed by our evaluation of $\gamma_T$.

\subsection{The zero temperature case}

\begin{figure}[tbh]
\vspace{0.5cm} %\epsfysize=5.5cm
\epsfig{figure=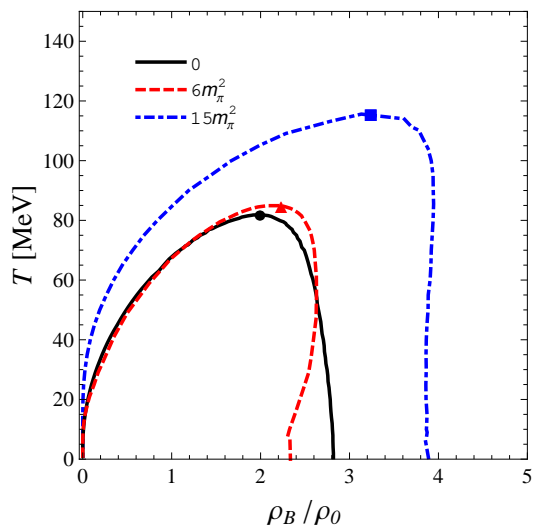,angle=0,width=7.cm}
\caption{  Phase coexistence boundaries in the $T-\rho_B$ plane ($\rho_B$ appears in units of the nuclear matter density, $\rho_0=0.17 \, {\rm fm}^{-3}$). The solid symbols indicate the location of the critical point for each value of $B$ which occur at ($T_c=81.1\, \MeV$, $\mu_c=324.7\,\MeV$) for $B=0$,
($T_c=84.9\, \MeV$, $\mu_c=324.7$) for $eB=6 \, m_\pi^2$, and ($T_c=115.8\, \MeV$, $\mu_c=279 \, \MeV$) for $eB=15\, m_\pi^2$. Taken from Ref. \cite {andre}.}
\label{fig1}
\end{figure}

\begin{figure}[tbh]
\vspace{0.5cm} %\epsfysize=5.5cm

\epsfig{figure=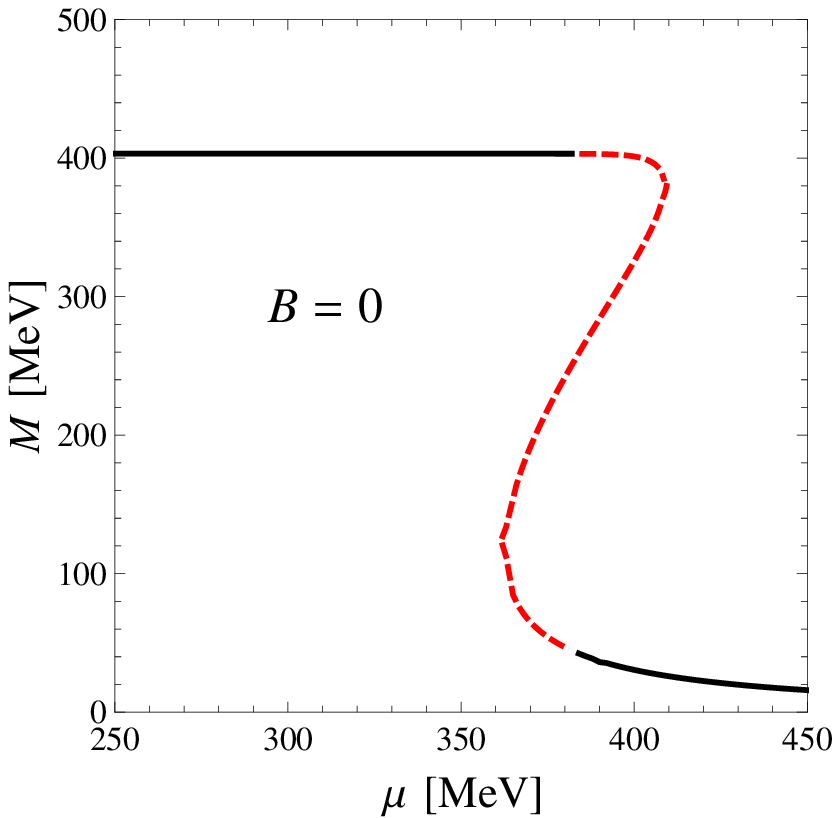,angle=0,width=5cm}
\epsfig{figure=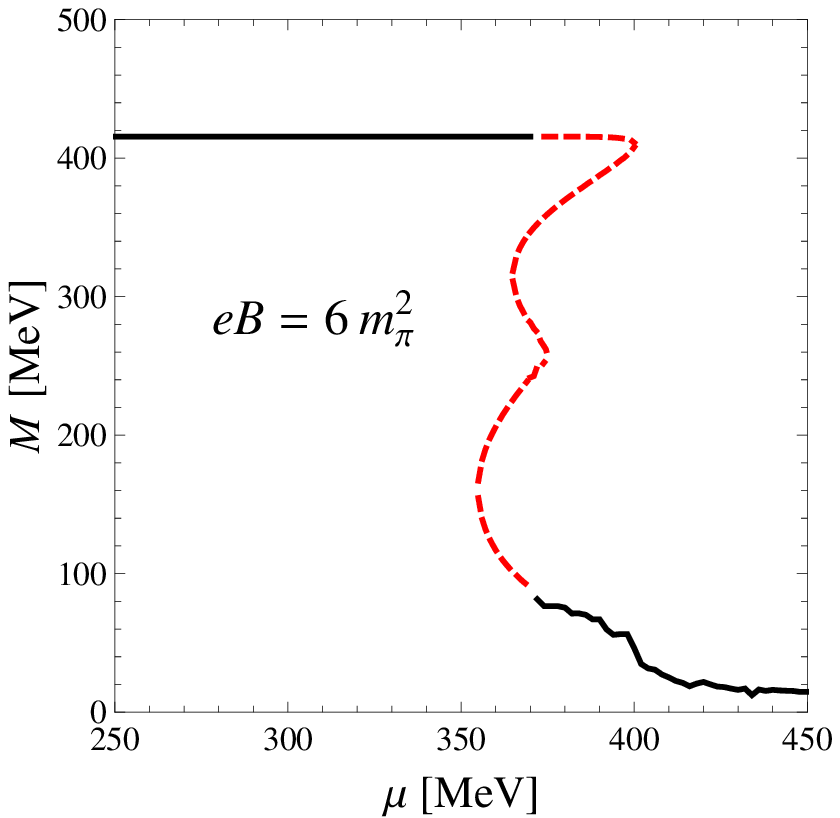,angle=0,width=5cm}
\epsfig{figure=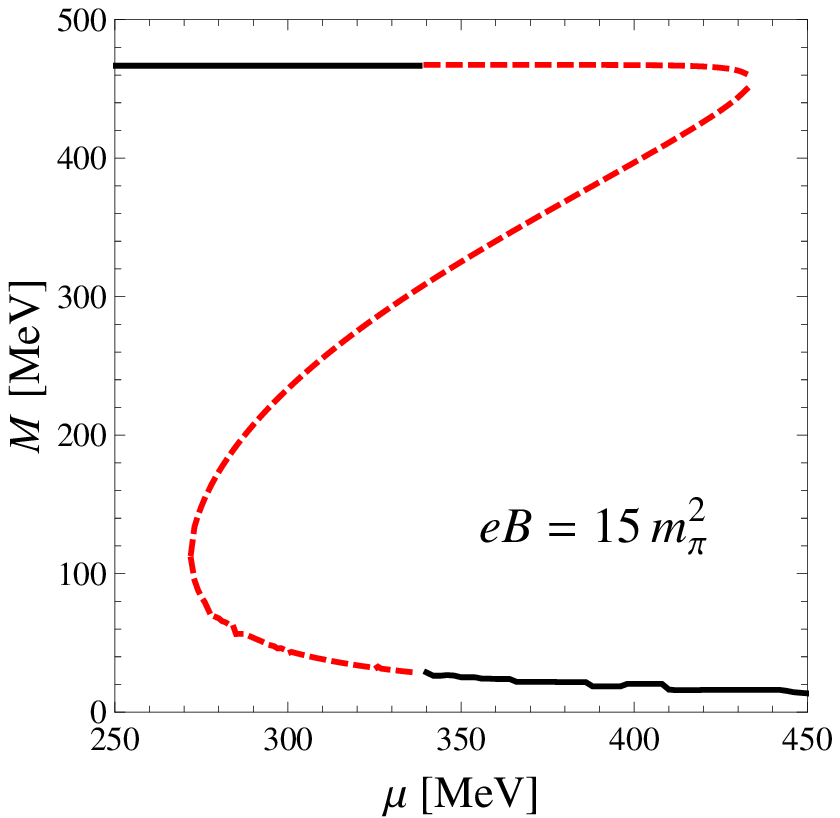,angle=0,width=5cm}

\caption{The quark effective mass, at $T=0$, as a function of $\mu$ for $B=0$ (left panel), $eB=6 m_\pi^2$ (center panel), and $eB=15 m_\pi^2$ (right panel). The continuous lines indicate the gap equation stable solutions and the dashed lines the unstable and metastable ones.   }
\label{fig2}
\end{figure}

\begin{figure}[tbh]
\vspace{0.5cm} %\epsfysize=5.5cm
\epsfig{figure=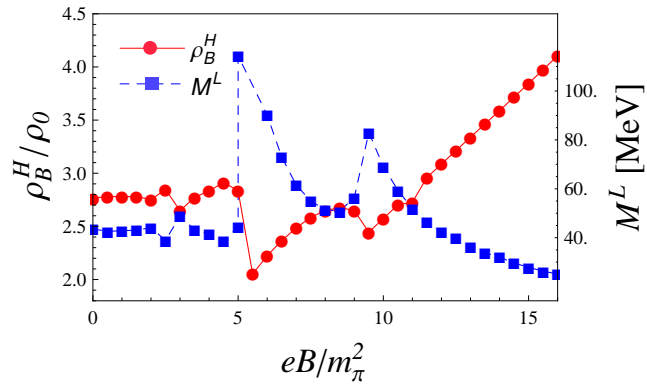,angle=0,width=8.5cm}
\caption{  The NJL model effective quark mass (squares) at the lowest value occurring at the transition, $M^L$, and the highest coexisting baryon density (dots), $\rho^H_B$ (in units of $\rho_0$), as functions of $eB/m_\pi^2$ at $T=0$. The lines are shown  just in order to guide the eye. Taken from Ref. \cite {andre}. }
\label{fig3}
\end{figure}

In order to illustrate how the method works and also to understand the type of oscillation displayed by Fig. \ref{fig1} it is convenient to concentrate in the $T=0$ limit since, in this case, the momentum integrals appearing in the thermodynamical potential can be performed producing equations which are easy to be analyzed from an analytical point of view.  Apart from that, this limit is very often considered in evaluations of the EoS for cold stars and it will be our starting point here. Then, in the next subsection we will analyze how the surface tension is influenced by  thermal effects.
At $T=0$ (and also at any other subcritical temperature) the grand potential can present multiple extrema representing
stable, metastable, and spinodally unstable matter in the neighborhood of the phase coexistence chemical potential and, as emphasized in Ref. \cite {mvj}, the extraction of  the surface tension by the
geometric approach requires the consideration of all these  extrema. In our case it is then important to know
all the gap equation solutions as displayed in Fig. (\ref {fig2}) which shows the effective quark mass, at $T=0$, for
$B=0$, $eB=6 m_\pi^2$, and $eB=15 m_\pi^2$. This effective mass is then used to determine the pressure from where all the other thermodynamic quantities, including the density, can be derived. In this figure, the continuous lines represent the stable solutions only and determine the Maxwell line which links the high effective mass value ($M^H$) to its low value ($M^L$) at the coexistence chemical potential where the phase transition occurs.  With these two stable solutions and upon using the Maxwell construction one obtains $f^M_T$. The dashed lines are obtained by considering the  unstable as well as the metastable gap equation solutions which lie within the spinodal region.  Considering all the gap equation solutions one then obtains $f_T(\rho)$ to determine the difference $\Delta f_T(\rho)$ which is the crucial ingredient in the surface tension evaluation. But before carrying out the evaluation let us discuss the origin of the the de Hass-van Alphen oscillations, for $\rho
 ^H$, which appear in Fig. \ref{fig1}  at $B\ne 0$.   Note from Fig. \ref {fig2}  that, at the coexistence chemical potential,  the gap equation for $eB=6m_\pi^2$, where the oscillations  are more pronounced, presents more solutions than the case $B=0$ or the case $eB=15 m_\pi^2$.
Then, the effective mass behavior displayed in Fig. \ref{fig2} allows us to understand the $\rho^H$ oscillations, shown in Fig. \ref {fig1}, by reviewing the discussion carried out in Ref. \cite {andre}. There it is shown that the decrease in $\rho^H$ for $eB=6m_{\pi}^2$, at low temperatures, can be understood in terms of the filling of the Landau levels. With this aim, we present Fig. \ref {fig3} which displays the  baryonic density and the  effective quark mass  as  functions of the magnetic field at $T=0$. To analyze the figure let us recall that, in the limit $T\rightarrow 0$, the baryonic density can be written\footnote {There is a misprint in Eq. (30) of Ref. \cite {prcsu2} where it should be $\rho_B$ instead of $\rho$.} as \cite {prcsu2}.
\begin{equation}
 \rho_B (\mu,B)= \theta (k_F^2)\sum_{f=u}^d \sum_{k=0}^{k_{f,max}} \alpha_k \frac{|q_f| B N_c}{6\pi^2}k_F \,\,,
\label{eq_rho_t0}
\end{equation}
where $k_F=\sqrt{\mu^2-2|q_f|kB-M^2}$ and
\begin{equation}
k_{f,max} = \frac{\mu^2 - M^2}{2|q_f|B} \,,
\end{equation}
or the nearest integer.
Eq. (\ref {eq_rho_t0}) shows that if $k_F^2 <0$ then $\rho_B=0$ which is precisely the low density value at $T=0$ which is easy to understand by recalling that the effective mass is double valued when the first order transition occurs presenting a high ($M^H$) and a low ($M^L$) value with  $M^L < M^H$ for $T<T_c$ and $M^L = M^H$ at $T=T_c$. Now, at $T=0$, $M^H$ corresponds to the value effective quark mass acquires when $T=0$ and $\mu=0$ (the vacuum mass) which corresponds to $M^H \simeq 403\, {\rm MeV}$ at $B=0$, $M^H \simeq 416\, {\rm MeV}$ at $eB=6 \, m_\pi^2$, and $M^H \simeq 467\, {\rm MeV}$ at $eB=15 \, m_\pi^2$. On the other hand, at $T=0$ the first order transition happens when  $\mu \simeq 383\, {\rm MeV}$ for $B=0$, $\mu \simeq 370 \, {\rm MeV}$ for $eB=6 \, m_\pi^2$ and
$\mu \simeq 339 \, {\rm MeV}$ for $eB=15 \, m_\pi^2$ so that $\rho^L=0$ even at the lowest Landau level (LLL), as required by $\theta(k_F^2)$ in Eq. (\ref {eq_rho_t0}). Then, to understand the oscillations let us concentrate on the $\rho^H$ branch which is shown, together with $M^L$ (the in-medium mass), in Fig. \ref{fig3}  where it is clear that both quantities have an opposite oscillatory behavior. The origin of the oscillations in these quantities can be traced back to the fact that  $k_{max}$ (the upper Landau level filled) decreases as the magnetic field increases. The first and second peaks, of the $M^L$ curve, correspond to the change from $k_{max}=1$ to $k_{max}=0$ for the \textit{up} and \textit{down} quark, respectively. For very low temperatures the value of $\mu$ at coexistence decreases with $B$ \cite {andre} so that, generally,  $k_{max}$ and $M$ must vary and when $k_{max}$ decreases, $M$ increases. It then  follows, from Eq. (\ref{eq_rho_t0}), that $\rho_B$ mu
 st decrease. When $k_{max}=0$ for both quark flavors there are no further changes in the upper Landau level and the low temperature oscillations stop at $eB \gtrsim 9.5 \,  m_\pi^2$.

\begin{figure}[tbh]
\vspace{0.5cm} %\epsfysize=5.5cm

\epsfig{figure=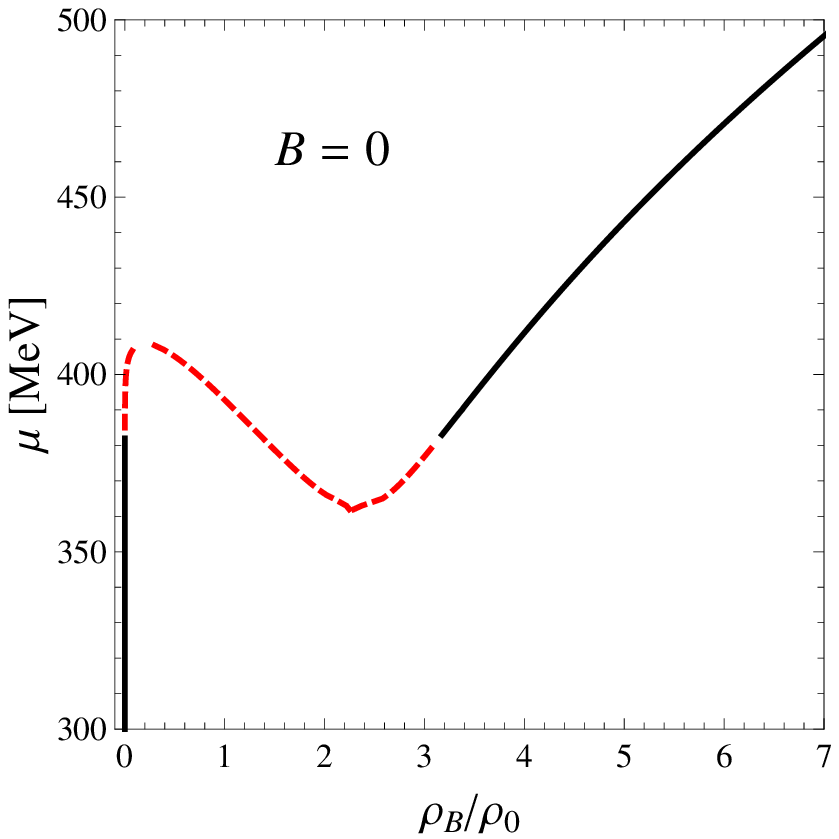,angle=0,width=5cm}
\epsfig{figure=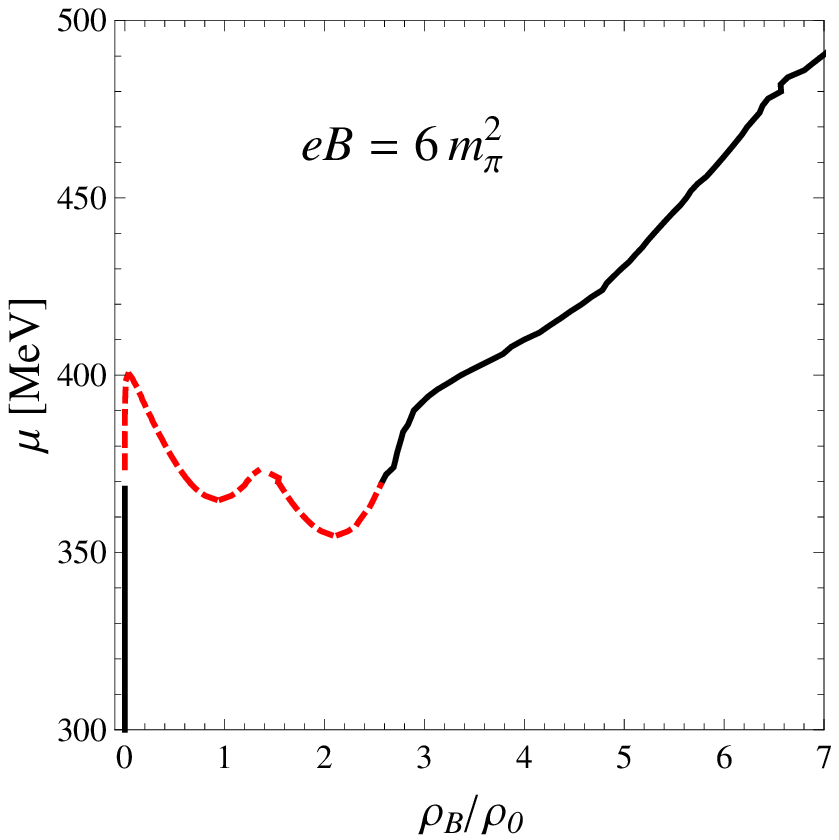,angle=0,width=5cm}
\epsfig{figure=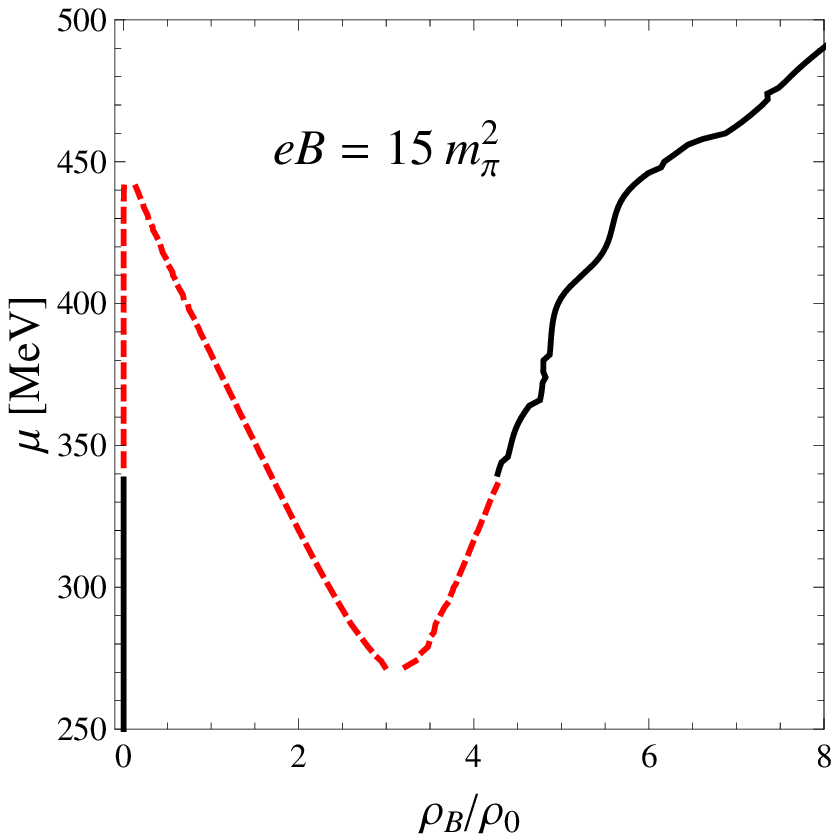,angle=0,width=5cm}

\caption{The chemical potential as a function of  of $\rho_B/\rho_0$ for $B=0$, $eB=6 m_\pi^2$, and $eB=15 m_\pi^2$. The continuous lines indicate the gap equation stable solutions and the dashed lines the unstable and metastable ones.    }
\label{fig4}
\end{figure}
\begin{figure}[tbh]
\vspace{0.5cm} %\epsfysize=5.5cm

\epsfig{figure=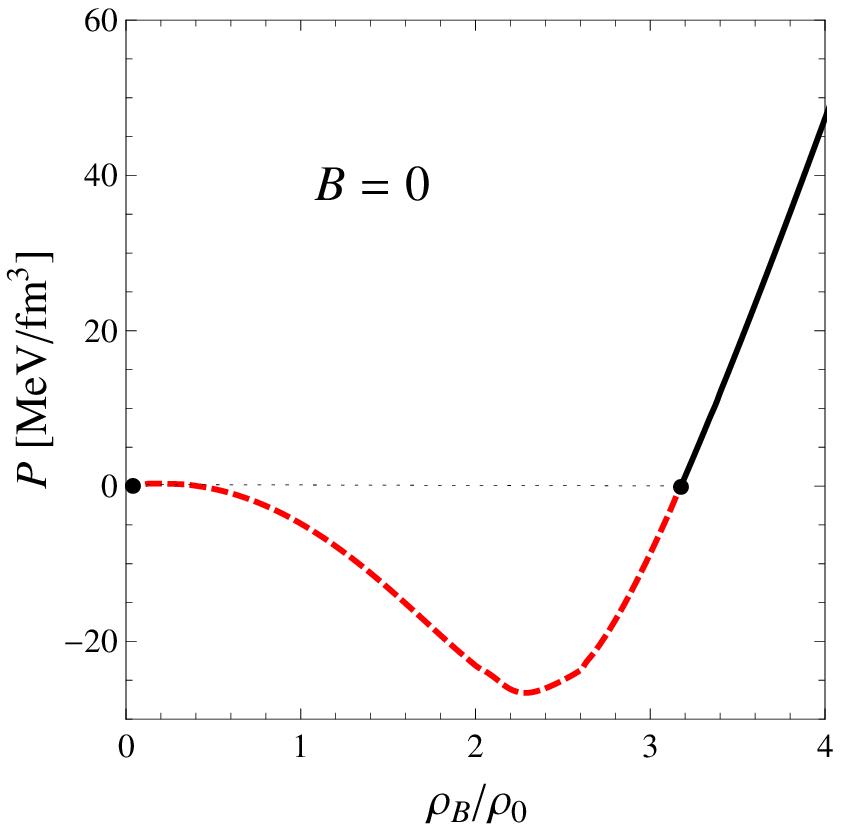,angle=0,width=5cm}
\epsfig{figure=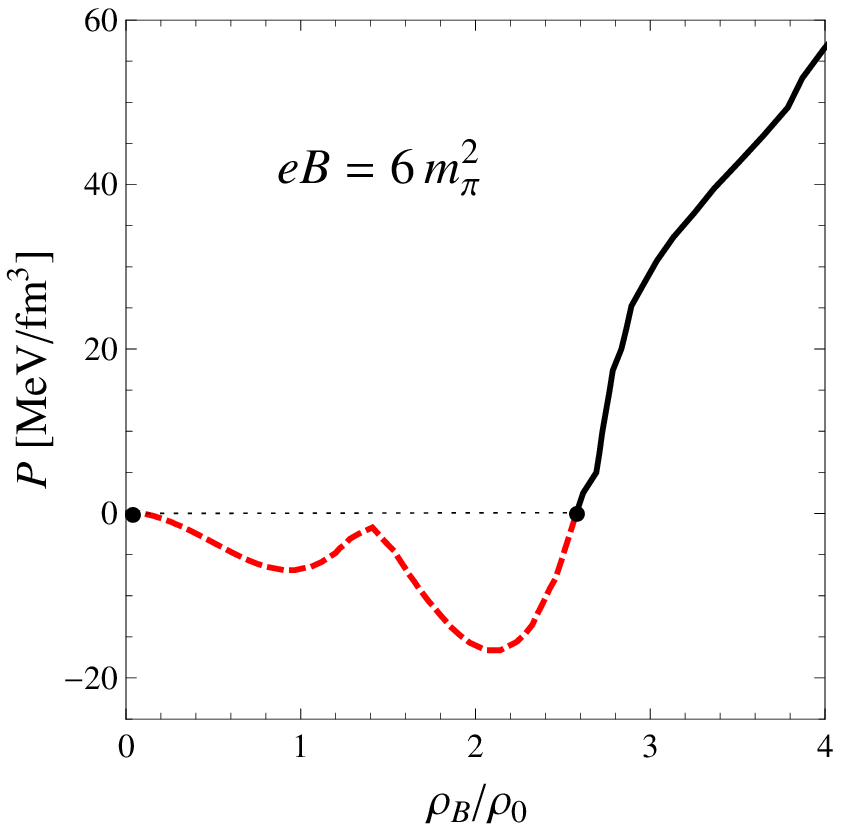,angle=0,width=5cm}
\epsfig{figure=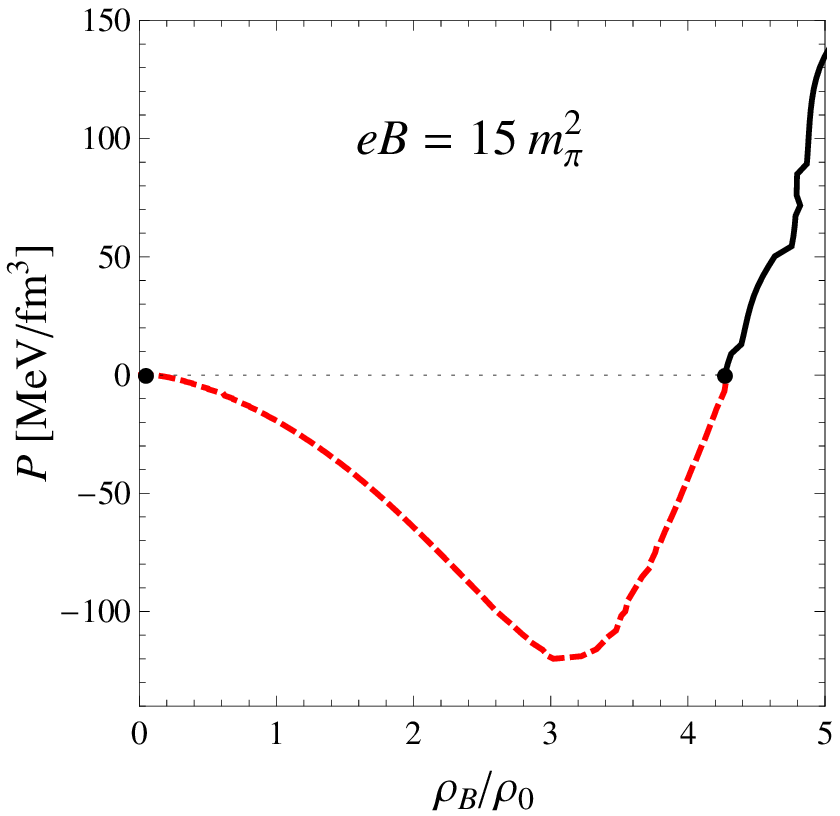,angle=0,width=5cm}

\caption{  The pressure as a function of  of $\rho_B/\rho_0$ for $B=0$, $eB=6 m_\pi^2$, and $eB=15 m_\pi^2$. The continuous lines indicate the gap equation stable solutions and the dashed lines the unstable and metastable ones. The dotted lines joining the thick dots represent the Maxwell construction.}
\label{fig5}
\end{figure}

\begin{figure}[tbh]
\vspace{0.5cm} %\epsfysize=5.5cm
\epsfig{figure=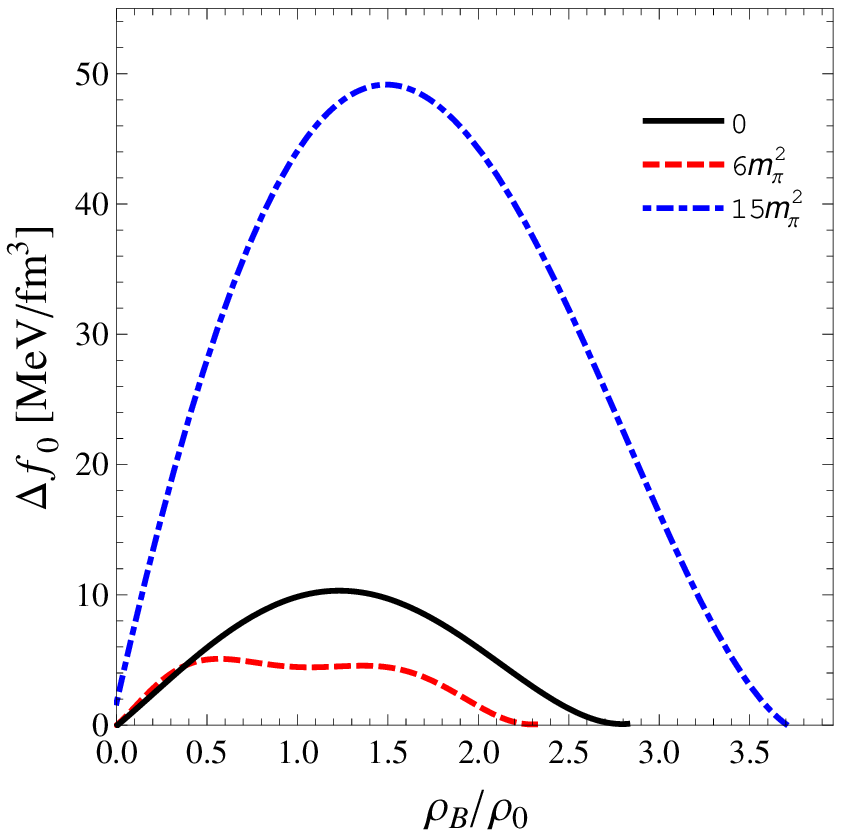,angle=0,width=7cm}

\caption{ The quantity $\Delta f_0$ as a function of $\rho_B/\rho_0$ for $B=0$, $eB=6 m_\pi^2$, and $eB=15 m_\pi^2$.}
\label{fig6}
\end{figure}

\begin{figure}[tbh]
\vspace{0.5cm} %\epsfysize=5.5cm
\epsfig{figure=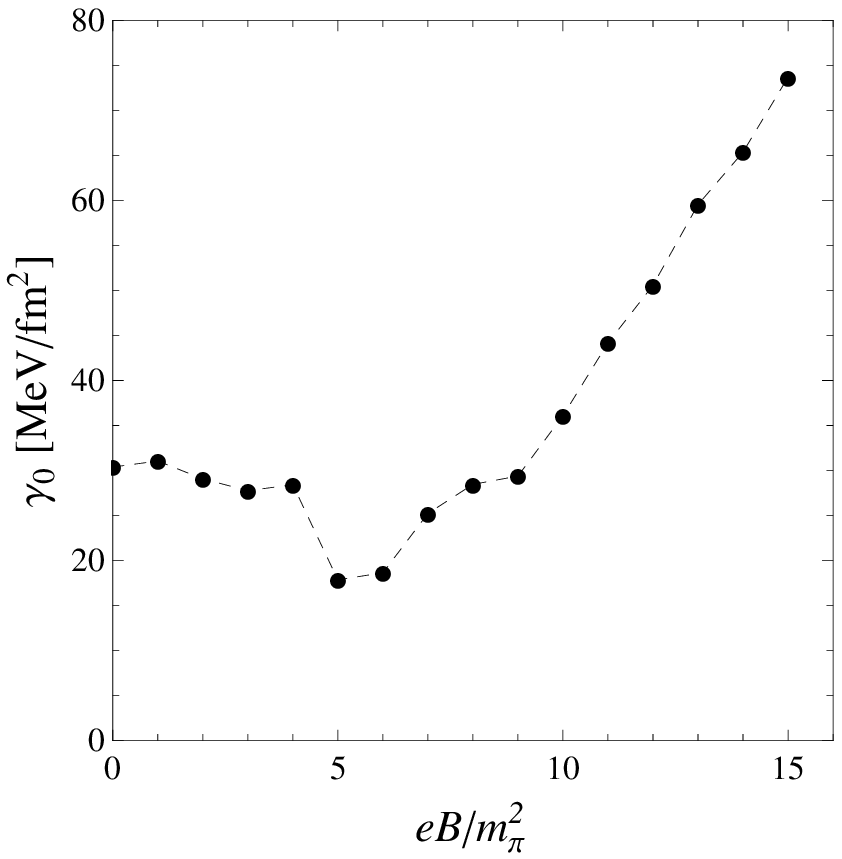,angle=0,width=7cm}

\caption{The surface tension at vanishing temperature, $\gamma_0$, as a function of $eB$ (in units of $m_\pi^2$). The lines are present just in order to guide the eye. }
\label{fig7}
\end{figure}

Let us now obtain the surface tension at vanishing temperature by first obtaining
 the difference between these two free energies,
$\Delta f_0(\rho)\equiv f_0(\rho)-f^M_0(\rho)$. Since $f_T(\rho)= \rho \mu(\rho) - P_T(\rho)$ one can start by evaluating $\mu(\rho)$ and $P(\rho)$ for uniform matter within the
thermodynamically unstable region of the phase diagram. Figs. \ref{fig4} and \ref{fig5} show the results for $\mu(\rho)$ and $P(\rho)$ respectively and, as before, the continuous lines reflect the stable gap equation solutions and the dashed lines the unstable and metastable ones.  It is then an easy task to obtain  a (positive)
deviation, $\Delta f_0(\rho)$, which determines the surface tension.
Fig. \ref {fig6} shows $\Delta f_0(\rho)$ for $B=0$, $eB=6 m_\pi^2$, and $eB=15 m_\pi^2$ displaying the expected oscillatory behavior around the $B=0$ case. Fig. \ref{fig7} shows the surface tension as a function of $eB$ at $T=0$ showing that it  oscillates  around the $B=0$ value  for $0< eB \lesssim 4 m_\pi^2$ before decreasing about $30\%$ for
$4 m_\pi^2 \lesssim eB \lesssim 6  m_\pi^2$. Then, after reaching a minimum at $eB \approx 6 m_\pi^2$ it  starts to increase continuously  reaching the $B=0$ value at $eB \approx 9 \; m_\pi^2$. After that, only the LLL is filled and $\gamma_0$ continues to grow with $B$.

Finally, table I summarizes all our results for $\gamma_0$, when $B=0$, $eB=6 m_\pi^2$, and $eB=15 m_\pi^2$, and also lists
the characteristic values ${\cal E}^g$, and $\rho^g$
as well as the location of the critical point $(T_c,\mu_c)$, and the upper integral limit (see Eq. (\ref{gamma})), $\rho^H$. For the present model approximation $\rho^L=0$ in all cases.
The table also shows that the values of the constituent quark mass,
at $T=0$ and $\mu=0$, grow with $B$ in accordance with the magnetic catalysis phenomenon.

\begin{table}[htb]	%	.........................................
\begin{center}
%\vspace{1cm}
\begin{tabular}{l||rlllllll}
%\hline \hline
$eB$ &
$\gamma_0~~ $ &  $M$ &
$T_c  $ &  $\mu_c  $&$\rho_B^H/\rho_0$ &$\rho^g_B/\rho_0$
& ~${\cal E}^g$  \\ \hline%\hline
0~            &  ~30.38   &
400 &81.1 & 324.7 &  2.73&2.03&495 \\ %\hline
 6         &     ~18.63  & 416   &
84.9 & 314.4 & 2.2 &  2.17 & 476  \\ %\hline
15 & ~73.68 & 467 & 115.8 & 279.0 & 3.8 & 3.17 & 705
\\ \hline
\end{tabular}
\end{center}
\caption{\label{tabfit} Summary of inputs and results at $T=0$ for different values of $eB$ (in units of $m_\pi^2$).
The length parameter was taken as $a=0.33\,\fm$.
The  characteristic energy
density ${\cal E}^g$ is given in $\MeV/\fm^3$,
and  the critical values $\mu_c$ and $T_c$ are given in $\MeV$.
The effective magnetic quark masses $M$ (at $\mu=0$) is also given in MeV while the resulting zero-temperature surface tension $\gamma_0$ is given in $\MeV/\fm^2$. In all cases $\rho_B^L=0$ and $\rho_0=0.17 {\rm fm}^3$. }
\end{table}

%------------------------------------------------------------------------
\subsection{Thermal effects}

\begin{figure}[tbh]
\vspace{0.5cm} %\epsfysize=5.5cm
\epsfig{figure=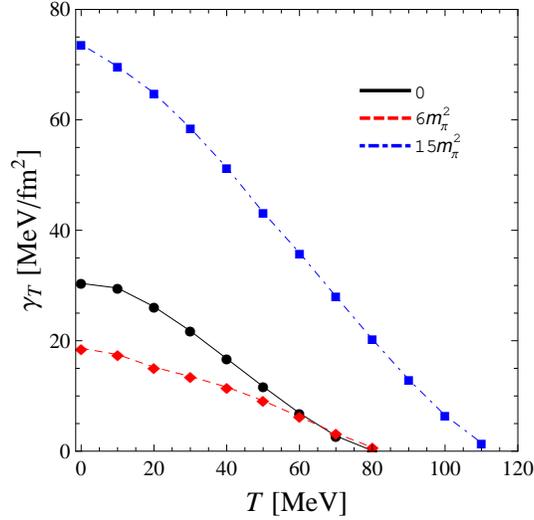,angle=0,width=7cm}
\caption{ The surface tension, $\gamma_T$, as a function of the temperature for for $B=0$, $eB=6 m_\pi^2$, and $eB=15 m_\pi^2$. The lines are present just in order to guide the eye. }
\label{fig8}
\end{figure}

Let us now investigate how thermal effects influence the
interface tension since this quantity is expected to decrease with increasing temperature
because both the coexistence densities
and the associated free energy densities
move closer together at higher $T$;
they ultimately coincide at $T_c$ where, therefore, the tension vanishes.
This general behavior is confirmed by our calculations,
as shown in Fig.\ \ref {fig8}.
The temperature dependence of the surface tension may be relevant for
the  thermal formation of quark droplets in cold hadronic matter
found in ``hot'' protoneutron stars whose temperatures, $T_*$,
are of the order 10--20 MeV \cite{bruno,hotstars,Olesen}.
For $T_*$ the relevant value of $\gamma_{T_*}$
may be estimated by using table I together with Fig.\ \ref {fig8}.
The temperature dependence of the surface tension is also
important in the context of heavy-ion collisions,
because it determines the favored size of the clumping
caused by the action of spinodal instabilities
as the expanding matter traverses the unstable phase-coexistence region \cite  {RandrupPRC79}.

\subsection{Other possible effects}

So far, our results for the surface tension were obtained within a certain model approximation, namely
 the standard two flavor NJL model at the mean field level. Therefore, one may wonder how other possibilities including  a different parametrization, strangeness, vector interactions, corrections beyond the MFA, and confinement, among others, would eventually influence our numerical predictions.
Let us start this  discussion with the parametrization issue in which case it becomes important to recall that, within the NJL model,  a stronger coupling increases the first order transition line in the $T-\mu$ plane.   This fact is reflected by  an increase of the coexistence region in the $T-\rho_B$ plane. Then,
 a stronger coupling should produce a higher surface tension which is  indeed the case, as  demonstrated in Ref. \cite {mvj} for $B=0$. For example, taking $\Lambda=631\, \MeV$, $G\Lambda^2=2.19$, and $m=5.5\, \MeV$ the critical point occurs at $T_c=46\, \MeV$ and $\mu_c=332\, \MeV$ while the effective quark mass value is $M=337 \, \MeV$ (compare with our values in table I). With this parametrization, one obtains $\gamma_0= 7.11 \MeV/\fm^2$ which is much smaller than our value, $\gamma_0= 30.38 \MeV/\fm^2$.  On the other hand, the surface tension value is expected to   increase by taking a higher coupling but  one should also remember that the effective quark mass  grows with $G$ and, with the set adopted here, we already have $M=400 \, \MeV$ which can be considered high enough\footnote{ In most works the coupling is chosen so that $M$ is about one third of the baryonic mass ($\approx 310 \, \MeV$).}. So, as far as the parametrization is concerned, our predictions could be lo
 wered by adopting coupling values which predict smaller values for the effective quark mass.

Next, let us point out that the presence of a repulsive vector channel may play an important role when treating the NJL at finite densities and, in this case, an interaction of the form $-G_V ({\bar \psi} \gamma^\mu \psi)^2$ is usually added to the Lagrangian density describing the model \cite {volker, buballa}. Then, regarding the  phase diagram, it has been established that the net effect of a repulsive vector contribution, parametrized by the coupling $G_V$,  is to add a term $-G_V \rho^2$ to the pressure weakening the first order transition \cite {FukushimaGV}. In this case,  the first order transition line shrinks, forcing the CP to appear at smaller temperatures,  while the first order transition occurs at higher coexistence chemical potential values as $G_V$ increases. In this case, the coexistence region decreases (this situation will not be affected by the presence of a magnetic field \cite {robson}) and should produce an even smaller value for the surface tension.

With respect to the MFA adopted here we believe that further improvements will only reduce the surface tension since evaluations performed with the nonperturbative Optimized Perturbation Theory (OPT), at $G_V=0$, have shown \cite {prc1} that already at the first non trivial order the free energy receives contributions from two loop terms which are $1/N_c$ suppressed.  It turns out that  these exchange (Fock) type of terms, which do not contribute at the large-$N_c$ (or MFA) level, produce a net effect similar to the one observed with the MFA at $G_V \ne 0$. This is due to the fact that the OPT pressure displays a term of the form $-G_S/(N_f N_c) \rho^2$ where $G_S$ is the usual scalar coupling so that a vector like contribution can be generated by quantum corrections even when $G_V=0$ at the Lagrangian (tree) level. The relation between the MFA (at $G_V \ne 0$) and the OPT (at $G_V=0$) and their consequences for the first order phase transition has been recently analyzed in g
 reat detail \cite {ijmpe}.  Based on this result one concludes that, in principle,  the inclusion of corrections beyond the mean field level  may contribute to further decrease the value of  $\gamma_T$.

In stellar modeling, the structure of the star depends on the
assumed EoS built with appropriate models while the true
ground state  of matter remains a source of speculation.
It  has been argued \cite{itoh} that
{\em strange quark matter} (SQM) is the true ground  state of all matter
and this hypothesis is known as the Bodmer-Witten conjecture.
Hence, the interior of neutron stars should be composed predominantly of
$u,d,s$ quarks (plus leptons if one wants to ensure charge neutrality).
 The question of how strangeness affects $\gamma_0$ was originally addressed in
 Ref. \cite {mvj} where the three flavor NJL was considered yielding the value $\gamma_0 = 20.42\, \MeV/\fm^2$ which is still within the lower end of estimated values. Moreover, in their application to the three flavor Polyakov quark meson model, the authors of Ref. \cite {Mintz} have confirmed that the presence of strangeness should not affect the surface tension in a drastic way. Another important issue,  tread in Ref. \cite {Mintz}, concerns confinement which has been considered by means of the Polyakov loop. Also, in this case the main outcome is that the surface tension value is not too much affected when the quark model is extended by the Polyakov loop.

Together, all these remarks indicate that our (low end) estimates for $\gamma_T$ are basically stable to the inclusion of more refinements (such as strangeness and confinement) and can even be further lowered (e.g., by going beyond the mean field  level and/or  by including a repulsive vector channel).

%========================================================================
\section{Conclusions}
\label{conclude}
In this work we have evaluated the surface tension related to the first-order
chiral phase transition for two flavor magnetized quark matter by considering the NJL model in the MFA. To obtain this quantity we have used the prescription presented in Ref. \cite {RandrupPRC79}
which is straightforward once the uniform-matter equation of state
is available for the unstable regions of the phase diagram.
The surface tension determined in the present fashion is entirely
consistent with the employed model,
including the approximations and parametrizations adopted.
In practice one only needs to consider {\it all} the solutions to the gap equation
(stable, metastable and unstable) when generating the corresponding EoS. This method was previously employed to obtain the surface tension for the NJL in the absence of magnetic fields yielding $\gamma_0 \lesssim 30 \MeV/\fm^2$ which lies within the low end of available estimates ($\gamma_0\approx10-300{\rm\ MeV/fm^2}$) and is in agreement with other recent predictions which employ effective quark models \cite{leticia,Mintz}. The importance of this result concerns, for example, the possibility of a mixed phase occurring in hybrid stars since the existence of such a phase is possible when the surface tension has a low value  \cite {veronica}.

Our results have shown that, when a magnetic field is present, the surface tension value  presents a small oscillation around the $B=0$ value,  for $0< eB \lesssim 4 m_\pi^2$. Then, it  decreases for
$4 m_\pi^2 \lesssim eB \lesssim 6  m_\pi^2$ reaching a minimum at $eB \approx 6 m_\pi^2$ where the value is about $30 \%$ smaller than the $B=0$ result. After this point it starts to increase continuously  reaching the $B=0$ value at $eB \approx 9 \; m_\pi^2$. This result allows to conclude that the existence of a mixed phase remains possible within this range of magnetic fields and can even be favored at the core of magnetars if $B \sim 1.8 \times 10^{19}\, G$ (or, equivalently, $eB \sim 6\, m_\pi^2$). At about twice this field intensity the surface tension starts to increase rapidly with the magnetic field disfavoring    the presence of a mixed phase within hybrid stars. The origin of this behavior can be traced back to the oscillations present in the coexistence region which is a quantity of central importance in the evaluation of $\gamma_T$. We have also shown how the temperature affects this quantity by decreasing its value towards zero which is achieved at $T=T_c$, as e
 xpected. Other issues such as strangeness, the presence of a repulsive vector interaction, confinement, corrections to the MFA, as well as different parametrizations have also been discussed. We have argued that our surface tension values, which already rank at the low end of the available wide range of predictions, will be little affected by strangeness and confinement and will be even lowered by the presence of a repulsive vector term and/or by the inclusion of corrections beyond the mean field level so that a mixed phase within hybrid stars will be further favored by these improvements. On the other hand, with the adopted  model, the surface tension value could grow if one chooses a parametrization with a coupling greater than ours which in turn would lead to very high effective quark masses.

\acknowledgments

AFG thanks Capes for a PhD scholarship and  MBP thanks CNPq for  partial support. This work has also received funding from Funda\c c\~{a}o de Amparo \`{a} Pesquisa e Inova\c c\~{a}o do Estado de Santa Catarina (FAPESC). The authors are grateful to Veronica Dexheimer for her comments and suggestions.

%========================================================================

\end{document}